\documentclass[12pt]{article}

\advance\voffset by -2.0cm \advance\hoffset by -1.25cm
\usepackage{amsmath}
\usepackage{amssymb}
\usepackage{amsthm}

\newcommand{\RR}{\mathbb{R}} 
\newcommand{\ZZ}{\mathbb{Z}} 

\DeclareMathOperator{\Tr}{Tr} 

\newcommand{\N}{\mathcal{N}} 

\newcommand{\be}{\begin{equation}}
\newcommand{\ee}{\end{equation}}

\newcommand{\MM}{\mathbb{M}} 

\def\be{\begin{equation}}
\def\ee{\end{equation}}

\date{}
\author{Roberto Volpato\\[8pt]
Max-Planck-Institut f\"ur Gravitationsphysik \\ Albert-Einstein-Institut \\ 14476 Potsdam, Golm (Germany)}

\begin{document}

\begin{flushright}AEI-2012-008\end{flushright}

\vskip 2cm

\begin{center}
{\LARGE Mathieu Moonshine and symmetries of K3 $\sigma$-models}
\vskip 20pt
Roberto Volpato\\[8pt]
{\it Max-Planck-Institut f\"ur Gravitationsphysik \\ Albert-Einstein-Institut \\ 14476 Potsdam, Golm (Germany)}
\end{center}
\vskip 20pt

\begin{abstract}
  A recent observation by Eguchi, Ooguri and Tachikawa (EOT) suggests a relationship between the largest Mathieu group $\MM_{24}$ and the elliptic genus of K3. This correspondence would be naturally explained by the existence of a non-linear $\sigma$-model on K3 with the Mathieu group as its group of symmetries. However, all possible symmetry groups of K3 models have been recently classified and none of them contains $\MM_{24}$. We review the evidence in favour of the EOT conjecture and discuss the open problems in its physical interpretation.
\end{abstract}





\section{Introduction}

Two dimensional conformal field theories with $\N=(4,4)$ supersymmetry and central charge $c=6$ have received considerable attention in the physics literature, in particular because of their role as internal CFTs in type II superstring compactifications that preserve $16$ supersymmetries \cite{EOTY,Aspinwall:1996mn,NahmWend}. It is believed \cite{NahmWend} that all such theories arise as supersymmetric non-linear $\sigma$-models whose target space $X$ is either a four dimensional torus $T^4$ or a K3 manifold.  For each of these two topologies, there is a whole moduli space of models, depending on the choice of a Ricci flat metric and a closed B-field on the target space. These two classes of models can be distinguished by a topological invariant, the elliptic genus:
$$ \phi_{X}(\tau,z):=\Tr_{RR}\Bigl(q^{L_0-\frac{c}{24}}\bar q^{\tilde L_0-\frac{\tilde c}{24}}y^{2J^3_0}(-1)^{2J^3_0+2\tilde J^3_0}\Bigr)\ ,\qquad\qquad q:=e^{2\pi i\tau}\ ,\quad y:=e^{2\pi i z}\ .
$$ Here, $L_0,\tilde L_0$ are Virasoro generators and $J_0^3$ and $\tilde J_0^3$ are the zero modes of the Cartan generators of a $su(2)_1^{left}\times su(2)_1^{right}$ current algebra contained in the $\N=(4,4)$ superconformal algebra at $c=6$. 

\medskip

The elliptic genus can be computed explicitly using the constraints coming from supersymmetry \cite{EOTY}.  First of all, $\phi_{X}$ depends holomorphically both on $\tau$ and $z$, because the only states giving a non-zero contribution saturate the (right-moving) unitarity bound, {\it i.e.} they have $\tilde L_0$-eigenvalue $\tilde h=\tilde c/24=1/4$.
Moreover, the spectrum of these models is invariant under the spectral flow isomorphism of the $\N=(4,4)$ superconformal algebra. This implies that the elliptic genus satisfies some quasi-periodicity conditions \cite{EOTY}
\begin{align}\label{eq:jactmn2}
 &\phi_X(\tau,z+ \ell \tau + \ell') = e^{-2 \pi i 
(\ell^2 \tau+ 2 \ell z)} \phi_X(\tau,z)  &&
\ell,\ell'\in \ZZ \ ,\end{align}
%
as well as the modular properties
\begin{align}
\label{eq:jactmn1}
 &\phi_X\Bigl(\frac{a \tau + b}{c \tau + d} , \frac{z}{c \tau + d}\Bigr) =
e^{ 2 \pi i \frac{c z^2}{c \tau + d} } \, \phi_X(\tau,z)
\qquad &&\begin{pmatrix} a & b \\ c & d \\ \end{pmatrix} \in SL(2,\ZZ) \ ,
\end{align} that follow from standard CFT arguments (see section 2).
 Formally, these properties imply that $\phi_{X}$ belongs to the one-dimensional space of Jacobi forms of weight zero and index one \cite{EichlerZagier}. The normalisation depends on the topology of the target space, but not on the metric and B-field, so that it can be computed in some specific model
$$ \phi_{K3}(\tau,z)=8\sum_{i=2}^4\frac{\vartheta_i(\tau,z)^2}{\vartheta_i(\tau,0)^2}\ ,\qquad\qquad\qquad\qquad \phi_{T^4}(\tau,z)=0\ ,
$$  where $\vartheta_1,\ldots,\vartheta_4$ are the classical Jacobi theta functions \cite{EichlerZagier}.

\medskip

Although the explicit expression for the elliptic genus of K3 has been known for more than 20 years, new surprising properties have been recently discovered, starting from the seminal work by Eguchi, Ooguri and Tachikawa (EOT) \cite{EOT}. The EOT observations establish a connection between the elliptic genus of K3 and the Mathieu group $\MM_{24}$, a simple subgroup of the group of permutations of $24$ objects. The first evidence for this conjectural relationship emerges when one expands the elliptic genus into (left) $\N=4$ characters 
$$ \phi_{K3}(\tau,z)=20\; ch_{\frac14,0}^{short}(\tau,z)-2\; ch_{\frac14,\frac12}^{short}(\tau,z)+\sum_{n=1}^\infty A_n \; ch_{\frac14+n,\frac12}^{long}(\tau,z)\ .
$$ Here, $ch_{h,\ell}(\tau,z)$ denotes the character of a unitary Ramond $\N=4$ representation whose lowest weight state has $L_0$-eigenvalue $h$ and $su(2)$ isospin $\ell$ \cite{EOTY}. Unitarity imposes a lower bound $h\ge 1/4$ on the conformal weight, which is saturated if and only if the representation is short (BPS). The coefficients $A_n$ are the multiplicities of the long $\N=4$ representations corresponding to $h=1/4+n$. It can be proved that all $A_n$ are positive even integers, and the first few values are
\be\label{An}
\begin{array}{|c|ccccccc}\hline
n & 1 & 2 & 3 & 4 & 5 & 6 & \ldots \\ \hline
A_n/2 & 45 & 231 & 770 & 2277 & 5796 & 13915 & \ldots\\
\hline
\end{array}
\ee By comparing these coefficients with the dimensions of the irreducible representations of $\MM_{24}$
\be\label{m24dim}\begin{split} 
\{1,\quad 23,\quad 45,\quad 231,\quad 252,\quad 253,\quad 483,\quad 770,\quad 990, \quad 1035,\quad 1265,\\ 1771,\quad 2024,\quad 2277,\quad 3312,\quad 3520,\quad 5313,\quad 5544,\quad 5796,\quad 10395\}\ ,
\end{split}\ee one can notice several numerical coincidences \cite{EOT}.
 In fact, $A_n/2$, for $n=1,\ldots,5$, correspond exactly to the dimensions of some irreducible representations of $\MM_{24}$, while $A_6$ can be decomposed into a simple sum of such dimensions as $13915=10395+3520$. This suggests that the space of states contributing to the elliptic genus (at least, the ones belonging to long $\N=4$ representations)  can be decomposed as
\be\label{decomp} \bigoplus_{n=1}^\infty \;R_{\frac14+n,\frac12}^{\,\N=4}\otimes H_{n}\ ,
\ee where $R_{h,\ell}^{\N=4}$ is an irreducible $\N=4$ representation and $H_{n}$ a (real) representation of $\MM_{24}$ with $\dim H_n=A_n$. In particular, from table \eqref{An} one can argue that
\begin{align*} &H_1={\bf 45}\oplus \overline{\bf 45} & &H_2={\bf 231}\oplus \overline{\bf 231}\\
&H_3={\bf 770}\oplus \overline{\bf 770} & &H_4={\bf 2277}\oplus\overline{\bf 2277}\\
&H_5=2\cdot {\bf 5796} & &H_6=2\cdot {\bf 10395}\oplus 2\cdot {\bf 3520}\ .
\end{align*}
For larger values of $n$, the number of possible decompositions of $A_n$ into sums of dimensions \eqref{m24dim} grows very quickly, and it is difficult to make a guess for the representation $H_n$ based only on its dimension. In subsequent works \cite{Cheng:2010pq,Gaberdiel:2010ch,Gaberdiel:2010ca,Eguchi:2010fg}, it has been shown that there is a unique meaningful decomposition of $A_n$ into $\MM_{24}$-representations, and highly non-trivial evidence in favour of the EOT conjecture has been given (see section 2).

\medskip

This relationship between the elliptic genus of K3 and the Mathieu group $\MM_{24}$ is very reminiscent of a famous series of conjectures known as Monster Moonshine (see \cite{Gannon} for a review).
Although there is little doubt now that some sort of `Mathieu Moonshine' exists, its interpretation  is still an open problem. By analogy with the Monster Moonshine case, one would expect that there exists a non-linear $\sigma$-model on K3 whose group of symmetries contains $\MM_{24}$. However, this possibility has been ruled out in \cite{Gaberdiel:2011fg}, where all possible groups of symmetries of such models have been classified. 

\medskip

The paper is organised as follows. In section 2, following \cite{Cheng:2010pq,Gaberdiel:2010ch,Gaberdiel:2010ca,Eguchi:2010fg}, we review the evidence in favour of EOT conjecture. In section 3, we describe the classification  of symmetries of K3 models given in \cite{Gaberdiel:2011fg}, and in section 4 we discuss the consequences for the Mathieu Moonshine.

\section{Twining genera}

In this section, we provide stronger consistency checks for the EOT conjecture. The first step is to notice that, if the conjecture holds, the elliptic genus can be expanded as
$$ \phi_{K3}(\tau,z)=\dim H_{00}\;ch_{\frac 1 4,0}^{short}(\tau,z)-\dim H_0\; ch_{\frac{1}{4},\frac{1}{2}}^{long}(\tau,z)+\sum_{n=1}^\infty \dim H_n\; ch_{\frac 1 4+n,\frac 1 2}^{long}(\tau,z)\ ,
$$ where $H_{00}$ and $H_0$ are, respectively, the $\bf{23}\oplus \bf{1}$ and $\bf{1}\oplus\bf{1}$ representations of $\MM_{24}$, $H_n$ are the $\MM_{24}$-representations in  Eq.\eqref{decomp} and
$$ ch_{\frac14,\frac12}^{long}(\tau,z):=\lim_{h\searrow \frac{1}{4}}ch_{h,\frac12}^{long}(\tau,z)=2ch_{\frac14,0}^{short}(\tau,z)+ch_{\frac{1}{4},\frac{1}{2}}^{short}(\tau,z)\ .
$$
Then, for each $g\in\MM_{24}$, we can define the \emph{twining genus} \cite{Cheng:2010pq,Gaberdiel:2010ch}
\be\label{twigendef} \phi_g(\tau,z):=\Tr_{H_{00}}(g)\;ch_{\frac 1 4,0}^{short}(\tau,z)-\Tr_{H_{0}}(g)\; ch_{\frac{1}{4},\frac{1}{2}}^{long}(\tau,z)+\sum_{n=1}^\infty \Tr_{H_{n}}(g)\; ch_{\frac 1 4+n,\frac 1 2}^{long}(\tau,z)\ .
\ee Note that $\phi_g$ only depends on the conjugacy class of $g$ and, under the assumption that all $\MM_{24}$-representations $H_n$ are real, the twining genera for charge conjugated elements $g$, $g^{-1}$ will be equal. This leaves $21$ independent twining genera $\phi_g$, including the elliptic genus, which corresponds to $g$ being identity.

If $g\in\MM_{24}$ is a symmetry in any CFT with elliptic genus $\phi_{K3}$, then $\phi_g$ corresponds to the following trace
\be\label{twigenus} \phi_g(\tau,z)=\Tr_{RR}\Bigl({\bf g}\,q^{L_0-\frac{c}{24}}\bar q^{\tilde L_0-\frac{\tilde c}{24}}y^{2J^3_0}(-1)^{2J^3_0+2\tilde J^3_0}\Bigr)\ .
\ee By a standard argument in CFT, a trace of the form \eqref{twigenus} can be computed by a path-integral on a torus describing a closed string loop. More precisely, the twining genus $\phi_g$ is obtained by requiring the fields in the path-integral to satisfy $g$-twisted periodicity conditions along a certain non-trivial cycle of the torus. The group of modular transformations of the torus that preserve such $g$-twisted periodicity conditions is  
$$ \Gamma_0(N):=\Bigl\{\begin{pmatrix}a & b\\ c& d\end{pmatrix}\in SL(2,\ZZ)\mid c\equiv 0\mod N\Bigr\}\subseteq SL(2,\ZZ)\ ,
$$ where $N$ is the order of $g$. Thus, the twining genus $\phi_g$, for $g$ of order $N$, is expected to have simple automorphic properties under $\Gamma_0(N)$ \cite{Gaberdiel:2010ca}
\be\label{modprop} \phi_g\Bigl(\frac{a\tau+b}{c\tau+d},\frac{z}{c\tau+d}\Bigr)=
e^{2\pi i\frac{cd}{Nh}}e^{2\pi i\frac{cz^2}{c\tau+d}}\phi_g(\tau,z)\ ,\qquad \begin{pmatrix}a & b\\c & d\end{pmatrix}\in\Gamma_0(N)\subseteq SL(2,\ZZ)\ .
\ee
Here, the factor $e^{2\pi i\frac{cz^2}{c\tau+d}}$ is a consequence of the non-trivial modular transformation of the operator $y^{2J_0^3}$ in \eqref{twigenus}, while the phase $e^{2\pi i\frac{cd}{Nh}}$, where $h$ is an integer that divides $\gcd(N,24)$, represents a (possibly trivial) multiplier system for $\Gamma_0(N)$. A non-trivial multiplier ($h>1$) can arise when $g$ acts asymmetrically on the left- and right-moving sectors of the theory and it represents a failure of the level matching condition in the $g$-twisted sector.  If we assume that $g$ preserves the spectral flow automorphism, then also the periodicity conditions \eqref{eq:jactmn2} are satisfied and $\phi_g$ is a Jacobi form of weight $0$ and index $1$ under $\Gamma_0(N)$. Note that,  for $N=1$, Eq.\eqref{modprop} reduces to \eqref{eq:jactmn1}. 

\medskip

The existence of $21$ twining genera $\phi_g$, satisfying the modular properties \eqref{modprop} and admitting an expansion \eqref{twigendef} in terms of $\MM_{24}$-representations $H_n$, represents a highly non-trivial test of the EOT conjecture. In \cite{Cheng:2010pq,Gaberdiel:2010ch}, some partial lists of twining genera with the correct modular properties have been provided and the $\MM_{24}$-representations $H_n$, up to $n=7$, have been found by trial and error. The complete list of twining genera has been first derived in \cite{Gaberdiel:2010ca} (and independently in \cite{{Eguchi:2010fg}}). Using the orthonormality properties of  finite group characters, this result provides a systematic way to check, for each $n$, if a representation $H_n$ matching \eqref{twigendef} exists and, in this case, to determine it uniquely.
At the moment, $H_n$ has been identified up to $n\sim 1000$. Since the spaces of Jacobi forms satisfying \eqref{modprop} are in general rather small (most of them have dimension less than $10$), the existence of $1000$ representations $H_n$ matching \eqref{twigendef} represents very convincing evidence for the conjecture.

\section{Symmetries of K3 $\sigma$-models}

The physical interpretation of the EOT observation is not yet clear. The most obvious explanation would be that $\MM_{24}$ is the group of symmetries of some non-linear $\sigma$-model on K3. A slightly weaker condition would be that, for each $g\in \MM_{24}$, there is a certain K3 model that has a symmetry with the same order as $g$ and reproducing the twining genus $\phi_g$.  Both these possibilities have been ruled out by the following theorem proved in \cite{Gaberdiel:2011fg}:

\smallskip

\noindent {\bf Theorem:}
{\it Let $G$ be the group of symmetries of a non-linear $\sigma$-model on $K3$ preserving 
the $\N=(4,4)$ superconformal algebra as well as the spectral flow operators. 
One of the following possibilities holds:
\begin{list}{{\rm (\roman{enumi})}}{\usecounter{enumi}}
\item $G=G'. G''$, where $G'$ is a subgroup of $\ZZ_2^{11}$, and $G''$ is a subgroup of  
$\MM_{24}$ with at least four orbits when acting as a permutation on
$\{1,\ldots,24\}$
\item $G=5^{1+2}.\ZZ_4$
\item $G=\ZZ_3^4.A_6$
\item $G=3^{1+4}.\ZZ_2.G''$, where $G''$ is either trivial, $\ZZ_2$, $\ZZ_2^2$ or $\ZZ_4$.
\end{list}}
\smallskip

\noindent (Here $p^{1+2n}$ denotes an extra special group of order $p^{1+2n}$,
 and $N.Q$ denotes a group $G$ for which $N$ is a normal subgroup such that 
 $G/N\cong Q$, see \cite{Atlas}).
 
 \medskip
 
The proof is based on general properties of non-linear $\sigma$-models of K3 (see \cite{Aspinwall:1996mn,NahmWend} for a review). Any such model contains $24$ R-R ground states. Four of them are contained in a $\N=(4,4)$-supermultiplet with $h=\tilde h=1/4$ and $\ell=\tilde\ell=1/2$, while the remaining $20$ belong to distinct short supermultiplets with $h=\tilde h=1/4$ and $\ell=\tilde\ell=0$. The electric-magnetic charges for these fields, carried by D-branes, form a $24$-dimensional even self-dual lattice $\Gamma^{4,20}$ with signature $(4,20)$. 

For any given non-linear $\sigma$-model on K3, we are interested in describing its group $G$ of symmetries that commute with the $\N=(4,4)$ algebra and with the spectral flow isomorphism. In fact, if a symmetry $g$ of order $N$ satisfies these conditions, then the twining genus $\phi_g$ is a Jacobi form for $\Gamma_0(N)$ and admits a decomposition as in \eqref{twigendef} in terms of representations of $G$. 

The symmetries in $G$ must act by automorphisms on the lattice $\Gamma^{4,20}$ of D-brane charges. The conditions that $g\in G$ preserves the $\N=(4,4)$ algebra and the spectral flow amount to requiring $g$ to fix the four R-R ground states in the $\ell=\tilde\ell=1/2$ supermultiplet.  These four R-R states can be identified with a four-dimensional positive definite subspace $\Pi$ in $\Gamma^{4,20}\otimes\RR$. Therefore, the action of $G$ on the D-brane lattice $\Gamma^{4,20}$ leaves the subspace $\Pi\subset \Gamma^{4,20}\otimes\RR$ point-wise fixed. A non-linear $\sigma$-model on K3 is uniquely determined by the choice of the subspace $\Pi\subset\Gamma^{4,20}\otimes \RR$ \cite{Aspinwall:1996mn,NahmWend}; based on this property, it can be shown that the group of symmetries $G$ is isomorphic to the group of automorphisms of $\Gamma^{4,20}$ fixing $\Pi$ point-wise \cite{Gaberdiel:2011fg}.  The rest of the proof consists in a classification of all such groups, a problem that has been solved using techniques from lattice and group theory.

\medskip

Beside the classification theorem above, this proof establishes a direct connection between the structure of the lattice of D-brane charges in a given K3 model and its group of symmetries $G$. This provides a useful tool for the analysis of symmetries of K3 models. There are several examples of non-linear $\sigma$-model that can be explicitly described in terms of a rational CFT (torus orbifolds, Gepner models, Landau-Ginzburg models, etc.) and, in general, this description is sufficient to reconstruct the lattice of D-brane charges in full detail in terms of boundary states. On the other hand, a complete analysis of the group of symmetries is a much more complicated task: methods based on physical or geometrical intuition entail a concrete risk of overlooking symmetries that are not `natural' in the chosen description of the model. The construction in \cite{Gaberdiel:2011fg} provides a systematic way to describe the full group $G$ of symmetries by studying the automorphisms of the D-brane lattice. As an application, this method has led to the discovery of a rather `exotic' symmetry that mixes the twisted and the untwisted sector in a certain $T^4/\ZZ_2$ torus orbifold \cite{new}.  The precise way this symmetry acts on the fields of the theory is currently under investigation.

\section{Discussion}

The theorem in section 3 implies that not all $g\in\MM_{24}$  are realised as symmetries in some non-linear $\sigma$-model on K3. More precisely, an element of a given $\MM_{24}$-conjugacy class is a symmetry for some model if and only if it has at least four orbits in the $24$-dimensional permutation representation of $\MM_{24}$. This condition excludes the five conjugacy classes 12B, 21A, 21B, 23A, 23B, corresponding to three distinct twining genera (21B and 23B are charge conjugated of 21A and 23A, respectively, so they give rise to the same genus). At the moment, there is no good argument explaining why these three twining genera have good modular properties. 

%

\medskip

Moreover, there are several K3 models whose group of symmetry $G$ is not a subgroup $\MM_{24}$. In these cases, each of the $G$-conjugacy classes corresponds to a twining genus which is a well-behaved Jacobi form and satisfy a decomposition analogous to \eqref{twigendef} in terms of representations of $G$. This suggests that the `moonshine' might be extended to a group larger than $\MM_{24}$. The obvious candidate is the Conway group $Co_1$ \cite{Atlas}, since it contains  all possible groups $G$ of symmetries as subgroups \cite{Gaberdiel:2011fg} (and it is conjectured to be the smallest group with this property). However, any attempt to find a decomposition of the elliptic genus into representations of $Co_1$ (or any other group larger than $\MM_{24}$) has been unsuccessful so far. 

\medskip

If the `moonshine' only works for the group $\MM_{24}$, it is natural ask whether the symmetries corresponding to $\MM_{24}$ elements have any special properties that characterise them among the generic symmetries of K3 models. There are some hints that this is the case. For example, in \cite{Cheng:2011ay} it is shown that only the twining genera from $\MM_{24}$ elements satisfy a certain genus 0 property, analogous to the McKay-Thompson series in the Monster Moonshine. The precise way these properties are related to the Mathieu Moonshine, however, is still far from being understood.

%
%

\end{document}